# *Combination of window-sliding and prediction range method based on LSTM model for predicting cryptocurrency*

*Paraphrase:*

*Some of the ideas comes from my master's dissertation in University of Southampton, which might be with some similarity in Turnitin. I have contacted my supervisor, he agreed to use this idea as it's not a formal publishment.*

一作：Yifan Yao

二作：Lina Wang

*Abstract:*

*The present study aims to establish the model of the cryptocurrency price trend based on financial theory using the LSTM model with multiple combinations between the window length and the predicting horizons, the random walk model is also applied with different parameter settings. The object of this dissertation is the cryptocurrency, primarily the Bitcoin and Ethereum, of which the prices exhibit high volatility. Quantitative analysis is adopted as the method of this dissertation. The research tool is python programming language. Tensorflow package is employed to model and analyze research topics. The results of this study show the limitations of the LSTM and Random walk model for price prediction while demonstrating the different characteristics of both models with different parameter settings, providing a balance between the model's accuracy and the model's practicality.*

*Keywords—cryptocurrency, long short-term memory neural network( LSTM), window-sliding, random walk model*

# 1.Introduction

In the literature of contemporary cryptocurrency analysis, the topic of general dynamics for digital currencies is a popularity. **[1]** The transaction amount of cryptocurrencies has skyrocketed in 2017 with the super exponential growth in the capital market.**[2]** However, the movement of cryptocurrencies exhibits the high volatility which adds more uncertainty in the transaction market. Most articles on cryptocurrency and machine learning focus on the problems of model prediction, [3][4] but many of them ignore the mathematical principles behind the model and with the regardless of the relationship between accuracy and parameter settings. This leads to some seemingly accurate models that are not generally practical. This article will explore the relationship between mathematical principles and model accuracy and discuss the essence through phenomena. Considering that there are two theories in the financial market, one is that the stock price is predictable[6], and the other is that the stock price is completely unpredictable[5], which indicates the price is a random walk, so the machine learning model described below (e.g., LSTM and RNN) will verify the predictable hypotheses, and random walk theory is also applied in this article, which will be researched based on the previous study. [7] [8] [9]The article will explore the model's performance based on these two algorithms.

The Long-Short-Term memory (LSTM) and recurrent neural network (RNN) are frequently applied in this field, which are preferred over the conventional multilayer perceptron.{10} Sean McNally, compared the RNN and the LSTM model used on bitcoin[11], for the RNN implementation part, the author first taken the temporal length window by the autocorrelation function . In the LSTM part, the previous research [12] has illustrated that compared with the RNN, the LSTM outperforms RNN and ARIMA at learning long term dependencies. The ARIMA(Autoregressive Integrated Moving Average ) model is a type of time series model which is often used in the price prediction [13][14].A model comparison table is as below[12]:

The results of different model performance

| Model Results | | | | | | |
|---|---|---|---|---|---|---|
| Model | Temporal _Length | Sensitivity | Specificity | Precision | Accuracy | RMSE |
| LSTM | 100 | 37% | 61.30% | 35.50% | 52.78% | 6.87% |
| RNN | 20 | 40.40% | 56.65% | 39.08% | 50.25% | 5.45% |
| ARIMA | 170 | 14.7 | 1 | 1 | 50.05% | 53.74% |

Table [1]

The table shows that the precision and accuracy do not have significant difference between the two models, both LSTM and RNN models are capable on the training data with LSTM is more applicable to the long-term dependencies. The Long-Short-Term memory (LSTM) and recurrent neural network (RNN) are frequently applied in this field. **[10]**

As for the multiple window length settings, **(15)**, it is applied different window sizes based on the LSTM model to capture better features of the equipment , which concluded that various time window sizes have the positive impacts for recognizing various temporal dependencies among features, while **(16)**, used 10 combinations of sliding windows with prediction ranges to fully explore the accuracy improvement possibility for deep learning and concluded that if the window length is small and the prediction range is far ahead simultaneously, the RMSE will become lower than the basic method.

To sum up, In accordance with volatility, a nonlinear model should be applied to this topic. Many scholars have compared the RNN and the LSTM. According to the results, the LSTM model outperforms RNN since it is more suitable for long-dependencies. Significantly, the window sliding method with the different prediction range variable should be applied in this article. Furthermore, the theory of random walks in cryptocurrency prices is also experimented with respect to the predictable hypothesis of price.

# 2.Methodology
## 2.1 Principle and introduction of LSTM model

### 2.11 Start from RNN

RNN represents the recurrent neural network, time is a significant impact factor for RNN. [17]The output comes out with each moment's input combined with the state of the current model. In the Figure [1], the output $h_t$ comes out with both the input $x_t$ and the hidden state from moment t-1, which is provided by the looped edge. Theoretically, the recurrent neural network can be capable of sequences of arbitrary length. However, in practice, the problem of gradient dissipation or explosion will happen during the optimization for the too long sequence. Furthermore, the dissipation of the gradient will make the weight of previous layer not updated during the forward propagation, on the contract, the gradient explosion will make training process unstable, thus, the model cannot obtain the optimal parameters.

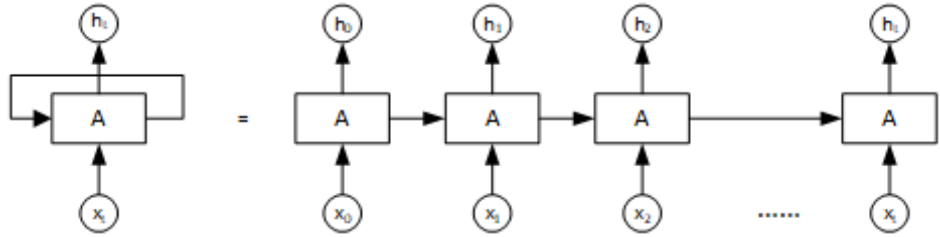
Figure[1] The structure of RNN

## 2.12 Mathematical Explanation of RNN

Given the 3 moments RNN unit, Figure. [2], assuming that the left input $S_0$ is a given value and no activation function exists in the neuron. Subsequently, the forward process is expressed as:

$$S_1 = W_x X_1 + W_s S_0 + b_1 \qquad O_1 = W_o S_1 + b_2$$
$$S_2 = W_x X_2 + W_s S_1 + b_1 \qquad O_2 = W_o S_2 + b_2$$
$$S_3 = W_x X_3 + W_s S_2 + b_1 \qquad O_3 = W_o S_3 + b_2$$

At the time of t=3, the loss function can be written as

$$L_3 = \frac{1}{2}(Y_3 - O_3)^2$$

RNN training is virtually to seek partial derivatives of $W_0, W_x, W_s, b_1, b_2$, adjusting them in order to obtain the minimum of $L_3$. According to the chain rule:

$$\frac{\delta L_3}{\delta W_0} = \frac{\delta L_3}{\delta O_3}\frac{\delta O_3}{\delta W_0}$$

$$\frac{\delta L_3}{\delta W_x} = \frac{\delta L_3}{\delta O_3}\frac{\delta O_3}{\delta S_3}\frac{\delta S_3}{\delta W_x} + \frac{\delta L_3}{\delta O_3}\frac{\delta O_3}{\delta S_3}\frac{\delta S_3}{\delta S_2}\frac{\delta S_2}{\delta W_x} + \frac{\delta L_3}{\delta O_3}\frac{\delta O_3}{\delta S_3}\frac{\delta S_3}{\delta S_2}\frac{\delta S_2}{\delta S_1}\frac{\delta S_1}{\delta W_x}$$

$$\frac{\delta L_3}{\delta W_x} = \frac{\delta L_3}{\delta O_3}\frac{\delta O_3}{\delta S_3}\frac{\delta S_3}{\delta W_s} + \frac{\delta L_3}{\delta O_3}\frac{\delta O_3}{\delta S_3}\frac{\delta S_3}{\delta S_2}\frac{\delta S_2}{\delta W_s} + \frac{\delta L_3}{\delta O_3}\frac{\delta O_3}{\delta S_3}\frac{\delta S_3}{\delta S_2}\frac{\delta S_2}{\delta S_1}\frac{\delta S_1}{\delta W_s}$$

Briefed as:

$$\frac{\delta L_3}{\delta W_x} = \sum_{k=0}^{t} \frac{\delta L_t}{\delta O_t} \frac{\delta O_t}{\delta S_t} (\prod_{j=k+1}^{t} \frac{\delta S_j}{\delta S_{j-1}}) \frac{\delta S_k}{\delta W_x}$$

$$\frac{\delta L_3}{\delta W_s} = \sum_{k=0}^{t} \frac{\delta L_t}{\delta O_t} \frac{\delta O_t}{\delta S_t} (\prod_{j=k+1}^{t} \frac{\delta S_j}{\delta S_{j-1}}) \frac{\delta S_k}{\delta W_s}$$

This formula suggests that the $\prod_{j=k+1}^{t} \frac{\delta S_j}{\delta S_{j-1}}$ part causes the gradient dissipation or explosion. With the activation function added, it is expressed as:

$$S_j = tanh(W_x X_j + W_x S_{j-1} + b_1 )$$

concluded that:

$$\prod_{j=k+1}^{t} \frac{\delta S_j}{\delta S_{j-1}} = \prod_{j=k+1}^{t} tanh' W_s$$

Where tanh derivative is always below 1. With the increase in $t$, the above formula's value turns closer to zero as long as $W_s$ is above 0 and below 1 as well, leading to the disappearance of the gradient. Subsequently, the above formula will become more and more infinite if $W_s$ is large, thus producing a gradient explosion, which explains why the LSTM is introduced.

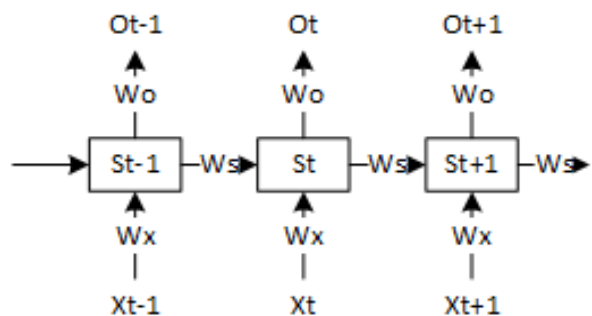

Figure[2] The inner structure of RNN

## 2.13 Mathematical Explanation of LSTM Model:

LSTM represents the Long-Short-Term memory, an RNN type. The $C_t$ is called current cell state which can be expressed as:

$$C_t = f_t \otimes C_{t-1} + i_t \otimes tanh(W_c[h_{t-1}, x_t] + b_c)$$

$f_t$ is called the forget gate, which can be expressed as:

$$f_t = \sigma(W_f[h_{t-1}, x_t] + b_f)$$

deciding which features can be employed for the calculation of $C_t$ from $C_{t-1}$. The current hidden output can be expressed as:

$$h_t = o_t \otimes tanh(c_t)$$

Besides, the input and output gates are expressed respectively as:

$$i_t = \sigma(W_i[h_{t-1}, x_t] + b_i)$$

$$o_t = \sigma(W_o[h_{t-1}, x_t] + b_o)$$

The above formulas show the activation function of 3 gates is sigmoid, revealing that the output of these three gates is either close to 0 or close to 1. This makes $\frac{\delta c_t}{\delta c_{t-1}} = f_t, \frac{\delta h_t}{\delta h_{t-1}} = o_t$ part is 0 or 1. When it is 1, the gradient can be transmitted well in the LSTM, significantly reducing the probability of the gradient dissipation. When the gate is 0, the information at the previous moment does not impacts the current moment indicating that there is not required to transmit the gradient goes back to update the parameters. [18]Accordingly, this explains the reason why the gradient can be solved using the LSTM model. See in Figure[3]

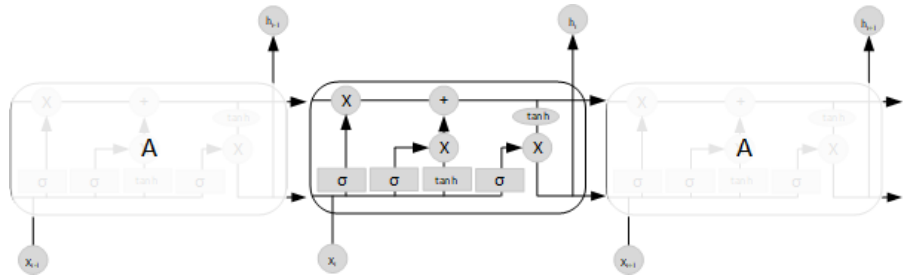

Figure[3] The structure of LSTM

## 2.14 Mathematical Explanation of Random Walk Model：

For the time series $\{x_t\}$, if it satisfies $x_t = x_{t-1} + w_t$, where $w_t$ denotes a white noise with a mean of 0 and a variance of $\sigma^2$, the sequence $\{x_t\}$ will be a random walk.[19] By definition, the $t$ at any $x_t$ moment refers to the sum of all historical white noise sequences that do not exceed the $t$ moment, so it is concluded that:

$$x_t = w_t + w_{t-1} + w_{t-2} + \cdots + w_0$$

The sequence mean and variance of random walk are presented as follows:

$$\mu_{x_t} = 0$$

$$var(x_t) = var(w_t) + var(w_{t-1}) + \cdots var(w_0)$$

$$= t \times var(w_t)$$

$$= t\sigma^2$$

Though the mean does not change with time $t$, due to the variance is the function that relates with the $t$, the random walk does not satisfy the stability. As time $t$ and the variance of $x_t$ are regulated, the stability is up-regulated. For the given interval $k$, the random walk covariance is performed as:

$$\text{Cov}(x_t, x_{t+k}) = \text{Cov}(x_t, x_t + w_{t+1} + \ldots + w_k)$$

$$= \text{Cov}(x_t, x_t) + \sum_{i=t+1}^{k} \text{Cov}(x_t, w_i)$$

$$= \text{Cov}(x_t, x_t) + 0$$

$$= t\sigma^2$$

From the concluded variance and covariance, the autocorrelation function $\rho_k$ (t) is calculated as below:

$$\rho_k(t) = \frac{\text{Cov}(x_t, x_{t+k})}{\sqrt{Var(x_t)}\sqrt{Var(x_{t+k})}}$$

$$= \frac{t\sigma^2}{\sqrt{t\sigma^2}\sqrt{(t+k)\sigma^2}}$$

$$= \frac{1}{\sqrt{1+k/t}}$$

Clearly, the autocorrelation function is related with the time $t$ and the interval $k$, indicating that if the random walk model has a long time series while the interval is quite small, the autocorrelation coefficient is approximated as 1. In other words, if there is a model predicting the stock price based on time t as the forecast for the $t+1$ value, the correlation coefficient between the actual value and the predicted value equals to the stock price sequence of $k=1$. In other words, the forecast of today's price as tomorrow's price is also very close to 1, which will mislead us that the model is accurate.

## 3. Data Collection

The data are all collected from the CoinMarketCap[20], which is authoritative website committed to cryptocurrency market value statistics. Only the Bitcoin and Ethereum data are adopted to train the LSTM model and the random walk model. The data column of both cryptocurrencies covering the open, high, low, close value daily, the transaction volume and the market capitalization. The raw ranges from April 2017 to December 2020 for nearly 3 years span. The training size parameter is 0.8, while the test size reaches 0.2.

## 4.Implementation

## 4.1 Training Process of Random Walk Model

## 4.1.1 Single Point Method Prediction

From the preliminaries illustrated below, the random walk model will learn the parameter $\sigma$, which is the only parameter of the random walk. The Figure[4] shows the model performance :

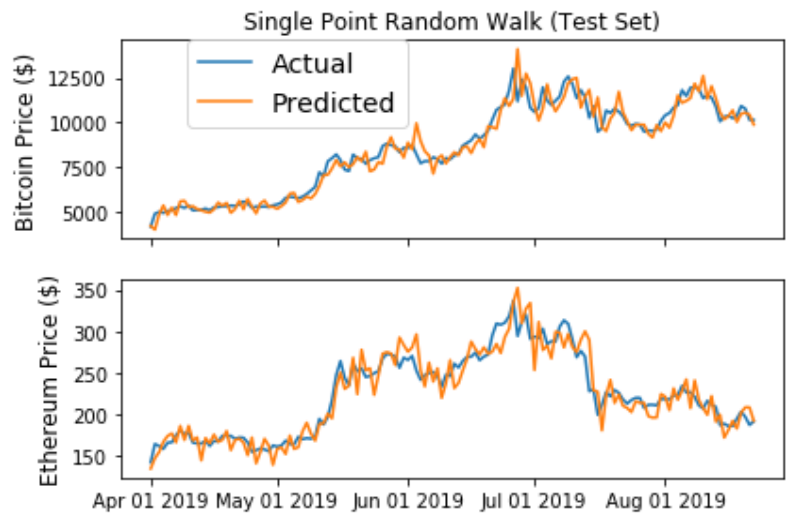

Figure[4] Singe point random walk model performace

Based on the preliminaries, the single point random walk model seems performing well, which is in accordance with expectation. The model just predicts the next day, so $k = 1$. Besides, the time span is 3 years, suggesting the $t$ is very large, so $\rho_k = 1$, implying that the forecast of next day is just the repeat of the current day, and due to the single point method selection, the error will reset every time which means every next input will all be the true data. Figure.[5][6] suggests that the prediction line is similar to the copy in the horizontal direction. The model seems accurate is attributed to the mathematical nature of random walk rather than the training process. Here, the model trained by the data in 2017 shows the details of the copy in the horizontal direction.

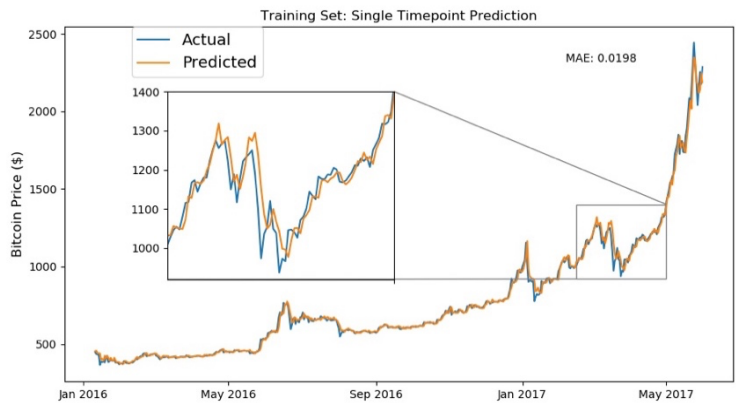

Figure.[5] The details of single point prediction on Bitcoin

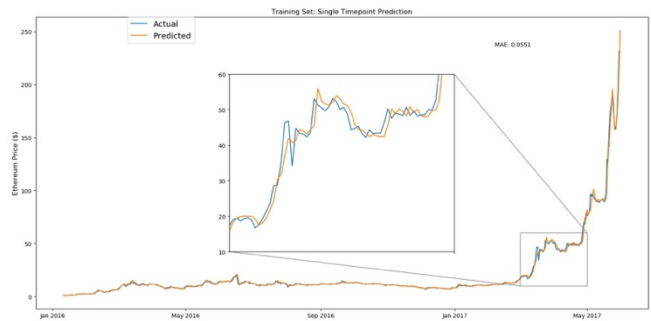

Figure.[6] The details of single point prediction on Ethereum

### 4.1.2 Multi-Point Method Prediction

As mentioned below, if the model intends to ignore the misleading accuracy caused by the nature of random walks, increasing the value of $k$ can solve this problem. That is to say, the interval of the random walk step will be larger instead of +1 days. Therefore, a multi - point prediction method is proposed. In such way, the error cannot be reset, which will be exacerbated by subsequent predictions. The training result can be seen below: Figure[7]

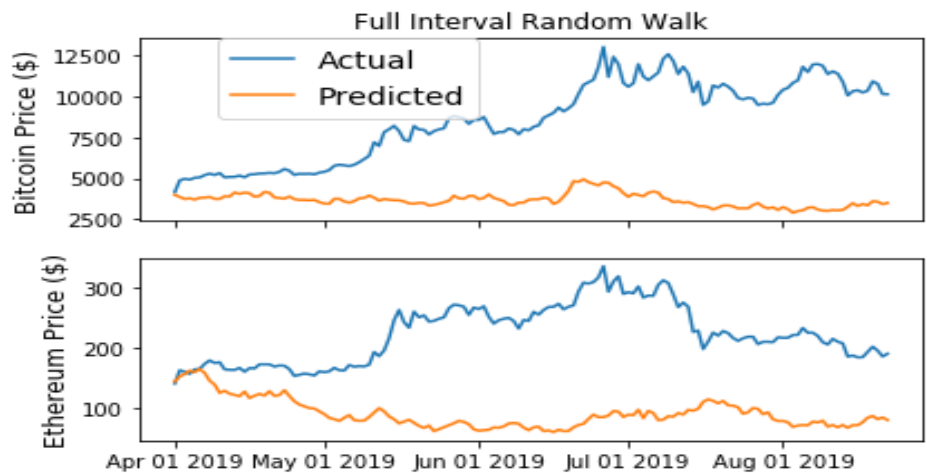

Figure[7] Full interval random walk performance

Obviously, changing the value of $k$ will cause a significant reduction in the model accuracy, $\rho_k(t)$ will not approach to 1 with the increase of $k$. That is to say, the result of the model is not associated with the nature of the Random Walk model. What is more, because the errors will be compounded by subsequent predictions, the predicting line is penalized seriously. What need to be noticed is that the Random Walk model is defined as $x_t = x_{t-1} + w_t$. That is, the price of the day is randomly changed based on the price of the previous day while the price difference is all included in the random item $w_t$. It can be seen from the above random walk model that the time series of the securities price will be in a random state and will not exhibit a certain observable or statistically determined trend. Compared with the machine learning model, the random walk model only explores the random item $w_t$, it does not learn from the inputs and learn any parameters or weights of the model. That is why whether the single-point model or the multi-point model are both not the ideal solution for predicting the trend of cryptocurrency.

## 4.2 Training Process of LSTM Model

### 4.2.1 Point to Point Method Prediction

The LSTM created is a 2-dimensional model using only the close price and the transaction volume features, considering the price of changes daily is an immense difference every period as the Figure.[8] shows below, which means the model will not converge, so the normalizing operation might be required.[21]

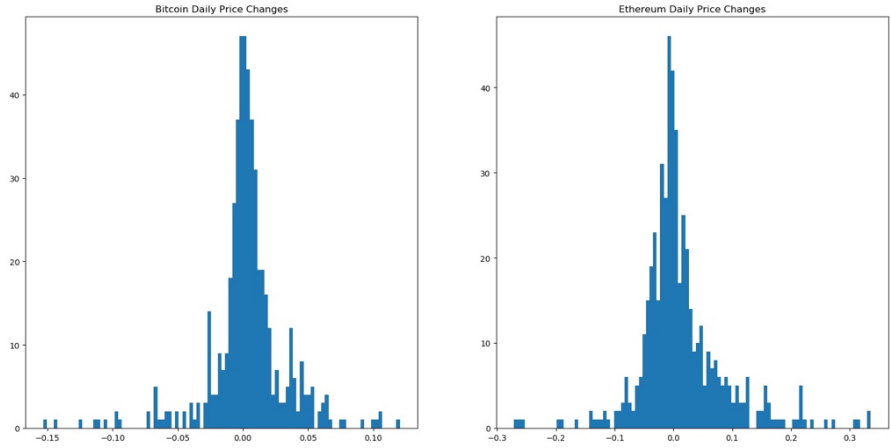

Figure[8] Daily price changes on Bitcoin and Ethereum

For the training data, to normalize the price changes, the following equation [1] is used, $p_i$ represents the current window price while the $p_0$ is the next window price. So the input and output will be a percentage format. For the test data, the output will be denormalized as a direct real price of prediction is expected to visualize, for the denormalization, the equation [2] will be used.

$$n_i = \left(\frac{p_i}{p_0} - 1\right) \qquad \text{Equation[1]}$$

$$p_i = p_0(n_i + 1) \qquad \text{Equation[2]}$$

Here the model uses MAE (Mean-Absolute-Error) equation to validate the error between the predicted value and the true value, which is the average of absolute errors that can better reflect the actual situation of the prediction value error.

$$\text{MAE} = \frac{1}{m}\sum_{i=1}^{m} |(y_i - \hat{y}_i)|$$

After the selection of parameters, the training dataset is used to train the model. The merge date starts from the 2017 to 2020 and the split size is 0.8, so the training dataset is mainly from 05-2017 to 10-2019. The Table[2] shows the Bitcoin training process of the model, it is obviously that from the epoch 18, the model started to converge as it lastly nearly stays at the MAE = 0.0330. And the Figure[13] shows the LSTM training process of the Bitcoin.

LSTM single point prediction training process

| epoch | 18/20 | 19/20 | 20/20 |
|---|---|---|---|
| Step-loss | 0.031 | 0.029 | 0.031 |
| mean_absolute_error | 0.031 | 0.029 | 0.031 |
| val_loss | 0.023 | 0.024 | 0.023 |
| val_mean_absolute_error | 0.023 | 0.024 | 0.023 |

Table[2]

After the convergence of the model, it is applied to the test dataset, which ranges from 11-2019 to 12-2020. The performance of the model on the Bitcoin test dataset is showed in the Figure[9]. As in the Figure. [12], both training set and the test set are all stop to decrease at epoch 20, after the epoch 20, the training set error will still go decrease, but the error on the test set will start to increase due to the model overfit problem.

The Figure[10][11] shows the performance of the model on the Ethereum.

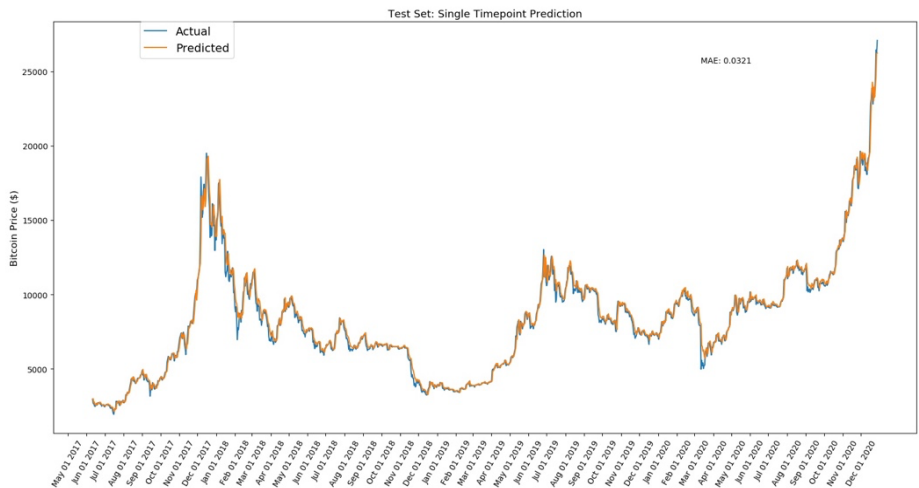

Figure.[9] The performance of LSTM with single point prediction on Bitcoin

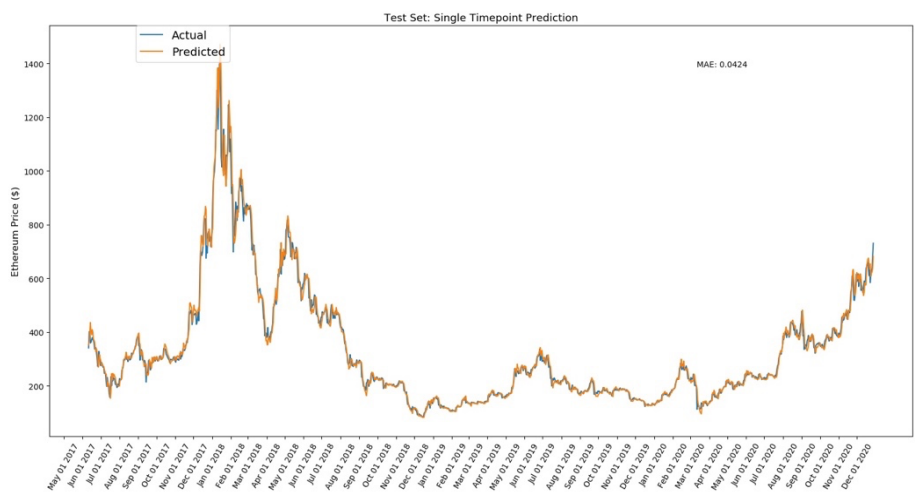

Figure[10]

The performance of LSTM with single point prediction on Ethereum

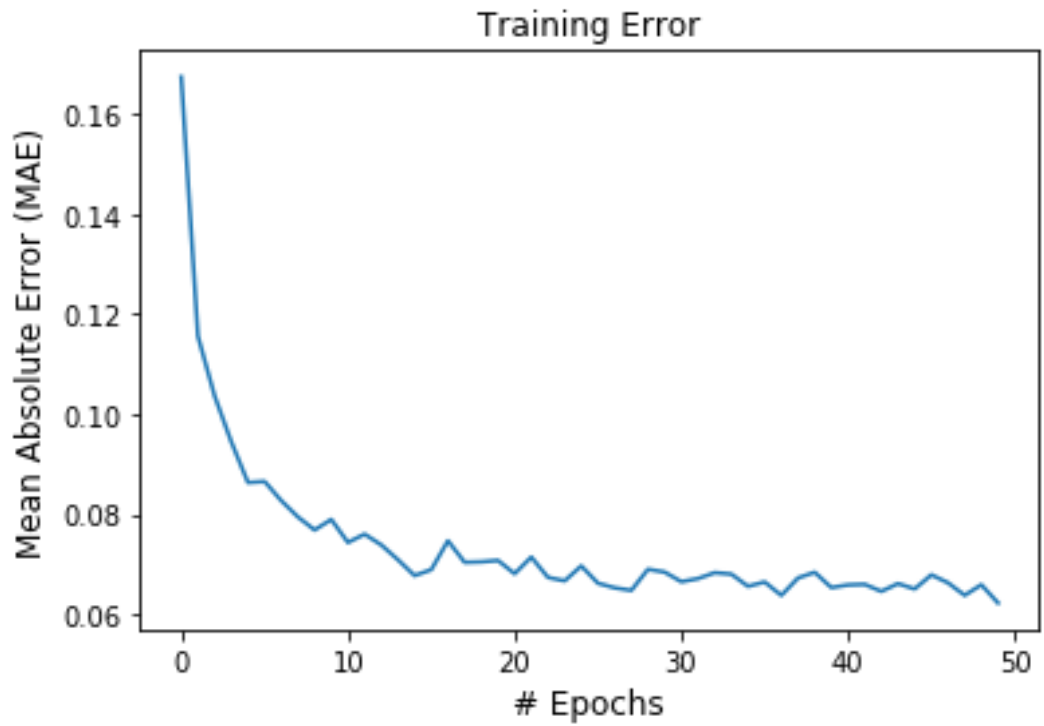

Figure.[11] The training error on Ethereum

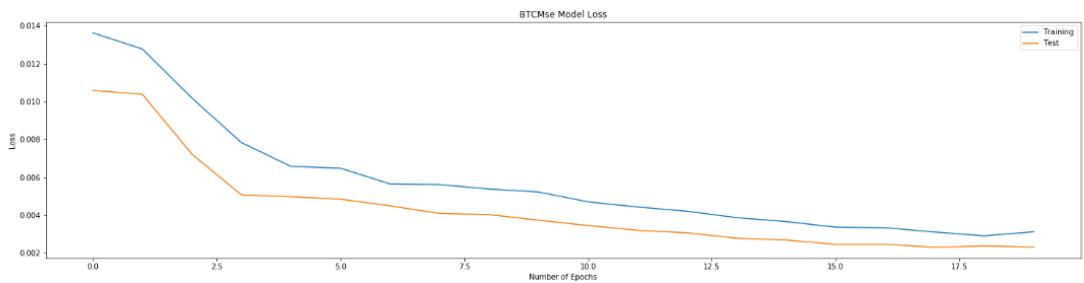

Figure[12] The model loss on Bitcoin

The model in this part used the point to point method. The point to point prediction is the process of making the model predict one single point value each time and plot the corresponding position in the figure, after predicting this point the window will slide to next point with the complete test data. Besides, the point to point method seems to be more accurate than the full interval prediction, whereas it does not imply that the point to point model outperforms the full interval model, since the error generated by each single prediction is reset each time, the neural network itself does not need to know the time series itself, all the inputs are based on the real value in every next prediction. For the ignorance of the errors, the model seems unsurprisingly accurate. Furthermore, in the Fig. [15], it is suggested that the predicted value is more like a horizontal translation of the true value. For instance, from the mid-May to mid-June 2019, several prices increased, and the peaks were following the fluctuations of the true values, which has an obvious hysteresis. In other words, the deep learning LSTM model regenerates an autoregressive model of order $p$, in these datasets area, the predicted value is the weighted sum of the previous $p$ values, as define below:

$$PredPrice = w_0 + w_1 * Price_{t-1} + \cdots w_p * Price_{t-p} + \epsilon_t, \epsilon_t \sim N(0, \sigma)$$

Where the next prediction will only be the true $Price_{t-p}$ value with the calculate weight because the point to point method will ignore the error of every previous prediction, which largely reduces the inaccuracy. Therefore, in order to maximize the advantages of LSTM based on time series and avoid the model updating the error at every step, the model will be improved from the following two indicators, the first is window length, and the second is the prediction range. Window length is the historical time used by LSTM, and the prediction range refers to the range of backward prediction by the data trained in window length. Figure[13]

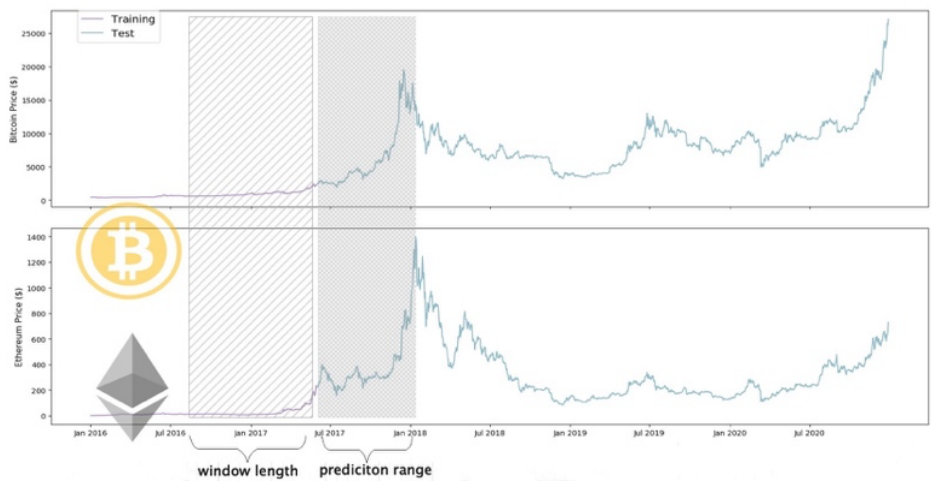

Figure[13] The explanation of window length and prediction range

## 4.2.2 Multi-Point Method Prediction with fixed window length selection

Unlike the limitation of point to point training method, the multiple timepoint method is more practical. Likewise, it initializes the test window and keeps moving to predict next point. Besides, it will move forward a full window size as well as resets the window with true test data while it moves to the $X_t$ point where the input window is already constituted by full past $X_{t-1}$ predictions. Thus, during the prediction, the error will not be fully reset, whereas the error will be accumulated in each full predicted window length, and the error will be reset again in a new window length. For this reason, the model will be more practical. It is neither as deceptive as single-point prediction, nor does it completely detour the model from the trajectory of the real point.

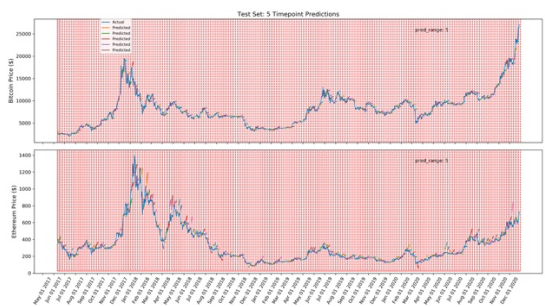

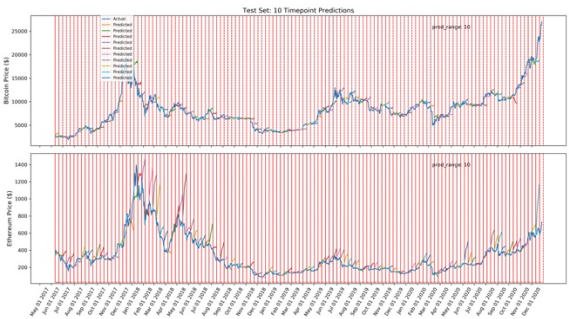

Figure.[14].                                    Figure.[15]

The model's performance with window length=10 prediction range = 5        The model's performance with window length=10 prediction range = 10

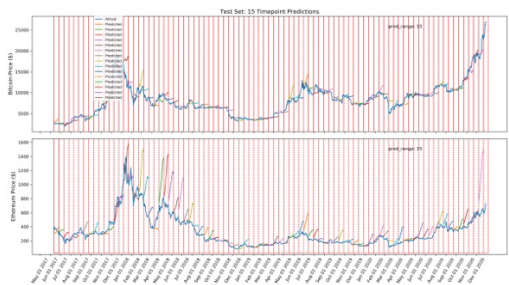

Figure[16] The model's performance with window length=10 prediction range = 15

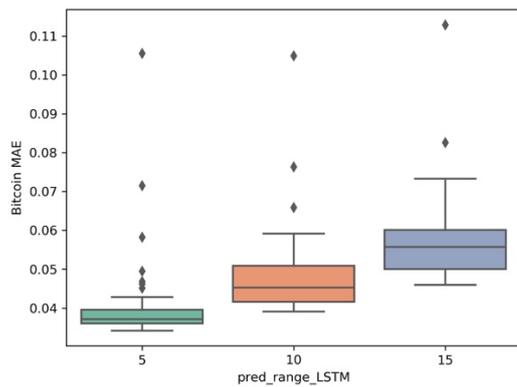 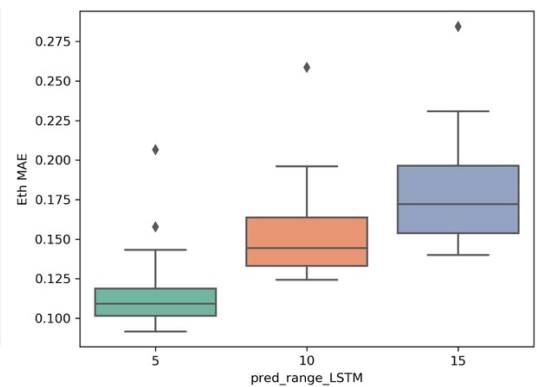

Figure[17] The box-plot of different prediction range with Bitcoin MAE  Figure.[18] The box-plot of different prediction range with Ethereum MAE

The Figure. [14][15][16] show that the multiple sequence LSTM does not perform well as expected. Besides, the red line in the figure is the prediction range. In the training process, the prediction range is set respectively at [5,10,15] while the window length is set at 10. The prediction of the model in each range does not reflect the price of the next trend, and the model seems to only predict the upward trend of the trend, while the price decline trend model does not seem to be aware. This may due to the selection of parameters or the selection of the length of the window, which reduces the model accuracy. In addition, the figure[17][18] point out that when the window length is fixed, as the number of prediction points increases, the MAE increases accordingly. Which indicates that in the condition of the same window length selection, the model will be more accurate with the less points amount.

## 4.2.3 Fixed Multi-Point Method Prediction with different window length selection

The previous part verifies the impact of different amount points selection on the model when the window length is fixed. This part will verify the impact of different window length on the accuracy of the model when the point amount is fixed. Similarly, the red line in the figure [19][20][21] refer to the different window length settings. In the training process, the window length is set respectively at [10,90,100] while the prediction range is set at 5. The figure[19] shows that at the condition of window_len=10, the model prediction trend performs similarly as the previous part, which seems to only predict the upward trend of the trend with the regardless of the decrease trend, while when the window_len= 50, the model could reflect the correct decrease trend generally and when the window_len = 90, the model could reflect all the trend but is not basically right, especially during the May 2019 -August 2019, the decrease trend of Bitcoin prediction is totally wrong.

From the figure[22][23], it could be concluded that with the fixed prediction range, the model accuracy decreases with the increase of the window length, which is caused by the accumulation of errors in the model.

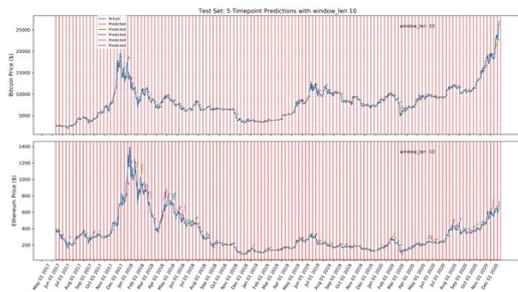 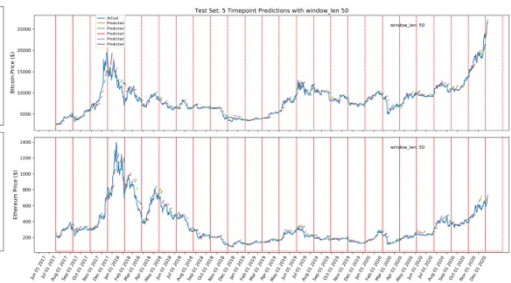

Figure.[19] The model's performance with window length=10  prediction range = 5

Figure.[20] The model's performance with window length=50 prediction range = 5

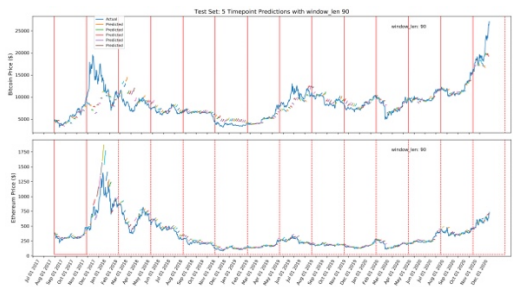

Figure.[21] The model's performance with window length=90  prediction range = 5

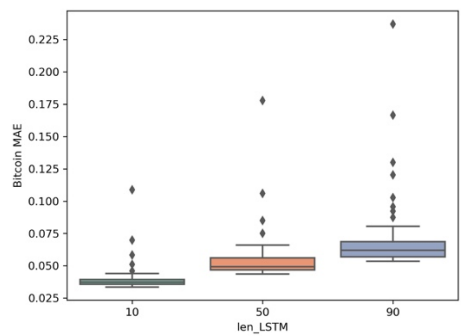 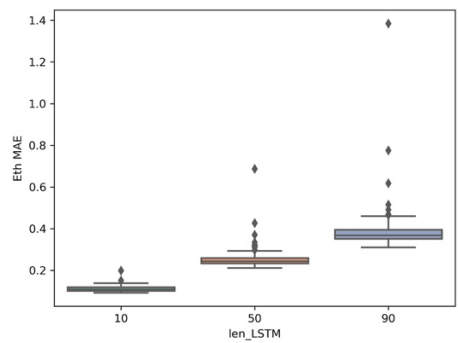

Figure.[22]   Figure.[23]

The box-plot of different window length with Bitcoin MAE.   The box-plot of different window length with Ethereum MAE

## Conclusion

According to the Table [3][4], after using different combinations of window length and prediction range it is found that when window length = 10 predict range = 5, the Bitcoin and Ethereum LSTM models reach the smallest errors, which are 0.037 and 0.113 respectively. Besides, it can be seen that although the single-point method has the smallest error, it is the result of the error being reset every time. However, in the real financial price market, only predicting the price trend of the next day is impractical. Thus, predicting the price over a period of time with a proper error reset frequency is more practical, that is, to have a certain prediction range. According to the Table [5], it shows the relationship between the interval of days and the accuracy. It can be seen that the error is not as large as expected. Therefore, it can be concluded that with a certain model accuracy guaranteed, the model has the best balance of practicality and accuracy based on the combination of window_length=10 and prediction range=5. In summary, it can be concluded from the study of Random Walk model and LSTM model that it is not appropriate to only focus on the model accuracy, considering the parameter setting and mathematical meaning as well as practicality also matters. Therefore, the balance between model practicality and accuracy is particularly important.

Limitation and future work
The range of window length settings is relatively large, the future research can be carried out within 10. Besides, the research objects are limited to Bitcoin and Ethereum, and more cryptocurrency can be introduced for experimental modeling.

The results of fixed window length with different prediction range

| Window _Length = 10 | | | | |
|---|---:|---:|---:|---:|
| Prediction_Range | 1 (single-point) | 5 | 10 | 15 |
| LSTM_MAE_Bitcoin | 0.032 | 0.037 | 0.045 | 0.057 |
| LSTM_MAE_Ethereum | 0.042 | 0.113 | 0.145 | 0.163 |

Table [3]

The results of fixed prediction range with different window length

| Prediction_Range = 5 | | | |
|---|---:|---:|---:|
| Window_Length | 10 | 50 | 90 |
| LSTM_MAE_Bitcoin | 0.037 | 0.051 | 0.063 |
| LSTM_MAE_Ethereum | 0.113 | 0.312 | 0.387 |

Table[4]

The error subtraction based on single day

| window_length=10 | | | |
|---|---:|---:|---:|
| Days_Interval (based on single day) | 4 | 9 | 14 |
| Error_Subtraction_Bitcoin | 0.005 | 0.013 | 0.025 |
| Error_Subtraction_Ethereum | 0.068 | 0.103 | 0.121 |

Table[5]